# Inferring Political Preferences from Twitter

Mohd. Zeeshan Ansari[1], Areesha Fatima Siddiqui[1*] and Mohammad Anas[1].

**Abstract** Sentiment analysis is the task of automatic analysis of opinions and emotions of users towards an entity or some aspect of that entity. Political Sentiment Analysis of social media helps the political strategists to scrutinize the performance of a party or candidate and improvise their weaknesses far before the actual elections. During the time of elections, the social networks get flooded with blogs, chats, debates and discussions about the prospects of political parties and politicians. The amount of data generated is much large to study, analyze and draw inferences using the latest techniques. Twitter is one of the most popular social media platforms enables us to perform domain-specific data preparation. In this work, we chose to identify the inclination of political opinions present in Tweets by modelling it as a text classification problem using classical machine learning. The tweets related to the Delhi Elections in 2020 are extracted and employed for the task. Among the several algorithms, we observe that Support Vector Machines portrays the best performance.

## 1 Introduction

Analyzing people's opinion on political views via surveys and polls is a time consuming and expensive task. It is impossible for humans to read and summarize all relevant documents and in terms of the expressed sentiments. Therefore, an automatic political sentiment analysis is required to deal with a large amount of text data [1]. People engaging in political issues actively are relying more on online sources to get informed about the latest news and events [2]. The use of social media such as Twitter has changed the way people express their views, feelings, opinions and this user-generated content in form of blogs, posts, tweets etc. are easily available publicly in an unstructured format. Twitter provides a platform to share and express political thoughts which have a huge impact on the political sphere in India. Twitter users mainly comprise of the educated class, political figures, prominent personalities which are influencers, are fewer in number when compared to the population of India. Their sentiments and opinion are considerably able to affect their follower's belief and opinions [21].

In this work, we queried the Twitter for scraping tweets that are related to Delhi Elections 2020 by using specific hashtags. We analyzed the tweets and labelled the polarity of sentiments of each tweet with respect to the major political parties taking part in elections. A total of 6060 tweets were collected and preprocessed into the relevant dataset including removal of all the noise that is present in the social media text. We carried out annotations on the extracted tweets by categorizing them into distinct sentiments classes as in *favour* or *against* a party or an alliance. Subsequently, we modelled the problem as a classical text classification problem and investigated the performances of several algorithms on the prepared dataset. The structure of the paper is as follows, Section 2 consists of the study of work that been has done in the past related to our topic of research. Section 3 elaborates the corpus acquisition method and theoretical dimensions to specify the adopted approach for our research. Section 4 and 5 consists of the procedure of experiments and the results obtained. Finally, in Section 6 we conclude the work done.

Mohd. Zeeshan Ansari
e-mail: mzansari@jmi.ac.in

Areesha Fatima Siddiqui
e-mail: areeshaf7@gmail.com

Mohammad Anas
e-mail: anas.1633.m@gmail.com

[1]Dept. of Computer Engineering, Jamia Millia Islamia, New Delhi 110025, India.

## 2  Related Work

There is a large increase in the availability of data in the form of text and documents that expresses the opinion of people which can be used for sentiment analysis and work in this field has grown rapidly [23]. The interest in analyzing a large amount of data produced around the world is generated due to the research on network and social media analysis and these data collected can be examined to discover interactions and behavioural patterns and have a better understanding of issues which are unrelated [3,22]. The sentiment in Twitter has been done in the domain of stock markets, politics and social movements for prediction and measurement [4, 5]. With the increase in the involvement of citizens in the electoral process empowers the democratic process on different levels and brings forward a new environment [6]. Prati and Hung (2017) [10] evaluated the exchanging of texts having a "defined ideological load", and the citizen contribution on Twitter during the Spanish electoral process in which they analyzed classes of segregation noticed in political orientations in messages that were posted digitally and they also considered event timelines [9, 14]. The volume of tweets on Twitter about the political affiliations was found out to be a good estimator in the 2009 German Elections while to detect the candidates in Singapore's 2011 national election by Twitter sentiment analysis was a failure [5, 11]. Colleoni et al. (2014) used classical learning and social media analysis approach to predict party preferences in an American Democratic and Republican voters database [15]. Several statistical approaches for democratic ideology estimation based on ideological stance and alignment of the traditional approaches to Twitter as a new source of data over the political campaign and keeping a track of people's behaviour and perception as the campaign developed over the time.

Sentiment classification is in a two-step task wherein the primary step, the data (tweets) relevant to our work is being collected and subsequently, in the second step, the sentiment from the data collected is being extracted. Relevant tweets contain words from a list of target keywords compiled either manually or semi-automatically from expanding a seed set [11]. Once the set of such messages has complied, several approaches are applied which are used to obtain the sentiment of the text. Unsupervised methods depend on a list of 'positive' and 'negative' keywords, which estimate a sentiment on the basis of the ratio of occurrences of keywords with respect to one another or just by counting the occurrence of each word concerning each other [5, 12]. Advanced approaches employ supervised learning techniques and train prediction models on either tweet classified manually or on the tweets which have an emotional context [8, 13, 16]. A. Jain and P. Dandannavar (2017) research study focuses on the combination of a lexicon-based method and a machine learning-based algorithm to define a mixed approach for performing sentiment analysis [17]. V. Sahayak et al. (2015) research studies intended to make it effortless for the companies to collect the feedback about the products they sell and for the customers who want to get other's opinions about a product prior to purchasing it [20]. Other research studies were focused on the conventional authorities of media, campaigns for elections, voter's social engagement and movement [18, 19].

## 3 Proposed Method

### 3.1 Political Corpus Preparation

The scraping of the politically inclined tweets from Twitter is carried using the automatic tool. The extraction of tweets is based on the hashtags and twitter account of popular politicians of respective political parties focused on Delhi Elections 2020. The hashtags that were used are: #DelhiElections2020 #DelhiPolls #delhielections etc. Total 6060 tweets related to Delhi Elections 2020 were obtained in the process for this task. The collection of tweets is significant to three major political parties which were mainly participating in Delhi Elections. We labelled the polarity of sentiment in tweets for every party and incorporated the inclination of a tweet towards or against the political party. All the tweets under consideration are updated and tweeted from Sept 2019 to January 2020.

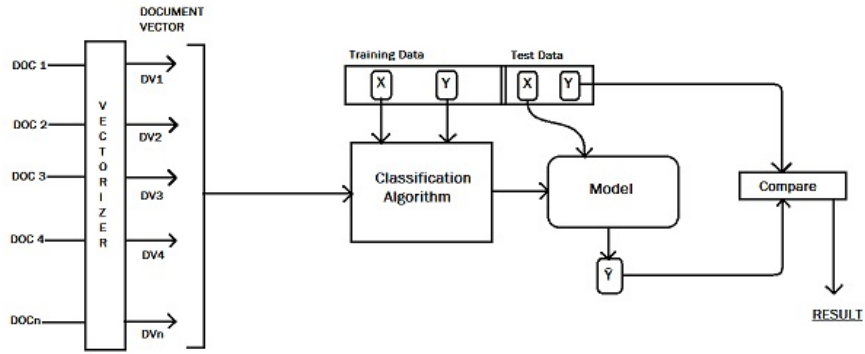

**Fig. 1.** The Text Classification Framework.

### 3.2 Preprocessing

Prior to analyzing the political inclination of sentiments in extracted tweets, we cleaned them into proper form for retrieving the relevant features required for successful classification. Several steps were taken to clean the tweets and remove the unwanted data such as (i) removing the username of the user who tweeted, (ii) all phone numbers, (iii) RT stands for retweets, (iv) all punctuation, (v) all trailing spaces, (vi) replacing every email id with "emailaddr", (vii) all break and white spaces with single space, (viii) every URL with "urladdr", (ix) all currency symbol with "moneysymb", (x) all numbers with "numbr" and (xi) converting all text into lowercase.

**Table 1.** Sentiment Classes.

| Class | Sentiment |
|-------|-----------|
| 1 | Support P1 |
| 2 | Oppose P1 |
| 3 | Support P2 |
| 4 | Oppose P2 |
| 5 | Support P3 |
| 6 | Oppose P3 |
| 7 | Non Relevant |
| 8 | None |

*3.3 Annotation*

After the preprocessing of tweets, annotation of tweets was performed in which each tweet was given a label according to its polarity of sentiment. Being considerably mindful of the electoral politics, the sentiment either positive or negative towards the respective political party or leader was identified for each tweet. We considered three political parties i.e. P1, P2 and P3 as primary parties because the majority of tweets were focused on these three parties for Delhi Elections 2020. The classes to which the cleaned tweets are labelled are present in Table 1.

Taking into consideration the approach similar to Wilson et al. (2005) we prepare the sentiment analysis model [7]. The tweets were labelled to eight different classes which are defined as follows: the tweet that supported and opposed party P1 was given class 1 and class 2 respectively. Similarly, classes 3 and 4 were given to tweets that supported and opposed party P2 and classes 5 and 6 were given to tweets that supported and opposed party P3. Instead of discarding the non-relevant annotations, they were assigned to a proper class (class 7) to analyze the dissemination of political and non-political tweets. The tweets that did not support or oppose any political party but were related to Delhi Elections were given class 8. There are some tweets which can belong to more than one class so for our annotation we labelled the tweet into the supporting sentiment taking the standard that support has a higher preference than opposing. The manually annotated corpus achieved the inter-annotator agreement over 97.3%. Each cleaned tweet is considered as a document for text classification. After using NGrams, the document vectors (DV1, DV2,…DVn) are formed. The vectorized data is then sent for training different classification algorithms such as Random Forest, Support Vector Machines, Logistic Regression and Naïve Bayes. After running classification algorithms on training data, models are built and their performance is compared.

The number of tweets extracted is 6060 and after categorizing the tweets into the respective classes according to their polarity of sentiment, the total number of tweets that belonged to each class is presented in Table 2. The analysis of annotated corpus shows that 17.06% of total tweets belong to a class that supports party P1, while 14.46% of tweets belong to a class that oppose party P1. The percentage of tweets that oppose party P2 is 13.04% and that support party P2 is drastically less percentage of 4.85%. The percentages of tweets that support and oppose party P3 are very less to a combined percentage of 1.48%.

Table 2. Corpus Statistics.

| Sentiment Classes | #tweets per class | % of tweets in each class |
|---|---|---|
| 1(Support P1) | 1034 | 17.06 |
| 2(Oppose P1) | 876 | 14.46 |
| 3(Support P2) | 294 | 4.85 |
| 4(Oppose P2) | 790 | 13.04 |
| 5(Support P3) | 37 | 0.61 |
| 6 (Oppose P3) | 53 | 0.87 |
| 7 (Non Relevant) | 623 | 10.29 |
| 8 (None) | 2353 | 38.82 |
| Total | 6060 | 100 |

## 4  Simulation and Experimental Results

### 4.1  Feature Extraction

N-gram language modelling is a significant probabilistic technique to extract essential features from a sample of text. We used unigrams, bigrams, trigrams and tetragrams of the tweets and created N-grams as presented in Table 3. The Term Frequency-Inverse Document Frequency being a significant method to convert a collection of raw documents to relevant features. It is denoted as tf-idf, tf-idf1, tf-ifd2, tf-idf3 respectively.

### 4.2  Classification Algorithms

The classification algorithms employed for this task encompass the different approaches to classification such as probabilistic models, ensemble methods, kernel methods etc.

*Support Vector Machine.* Support Vector Machine was first put forward by Cortis and Vapnik in 1992. It is a supervised learning model that can be used for classification, regression and outlier detection. Support vector machine is highly preferred as it has a simple structure and produces results with significant accuracy and less computation power. SVM also does not require a large number of features.

*Random Forest.* A Random Forest is a classification model that consists of a collection of tree-structured classifiers. It is an ensemble learning method that operates by constructing multiple decision trees at training time and giving the output as the class that is the most suitable for input x.

*Logistic Regression.* Logistic Regression is widely used to examine and define a relationship between a response variable, which is binary in nature, and a set of predictor variables. Logistic Regression fits data to a logit function and predicts the probability of occurrence of an event; therefore it is a type of predictive analysis. It is used for classification problems.

*Naïve Bayes.* Naïve Bayes is a collection of probabilistic classifiers that are based on Bayes theorem. In Naïve Bayes, every pair of feature classified is independent of each other and gives an equal contribution to the outcome. It is used for the classification problem. It requires fewer amounts of training data to estimate the necessary parameters.

**Table 3.** Count of N-grams.

| N-grams | Total number of N-grams |
|---|---|
| Unigrams | 6081 |
| Bigrams | 22468 |
| Trigrams | 42879 |
| Tetragrams | 50236 |

Table 4 shows a comparison of performance for different classifiers Random Forest, Naïve Bayes, Support Vector Machine and Logistic Regression with respect to different features like bigrams, trigrams and tetragrams. The performance of Support Vector Machine is best when compared to other classifiers in unigram. Support Vector Machine with tf-idf-unigram exhibit a precision of .85 while with tf-idf-bigram, the precision is 0.83. The precision of SVM with tf-idf-trigram and tf-idf-tetragram is 0.82. The recall and F1-score of Support Vector Machine are highest compared to others which is equal to 0.85 in tf-idf-unigrams. The performance of Naïve Bayes classifier is worst when compared to all classifiers. The precision for Naïve Bayes is least for unigram which is 0.79. Precision for Naïve Bayes is better than the precision of Logistic Regression for tetragram model whereas the same for trigrams.

**Table 4.** Performance of classifiers over various features

| | N=1 | | | N=2 | | | N=3 | | | N=4 | | |
|---|---|---|---|---|---|---|---|---|---|---|---|---|
| | P | R | F | P | R | F | P | R | F | P | R | F |
| Random Forest | 0.84 | 0.83 | 0.83 | 0.81 | 0.79 | 0.79 | 0.80 | 0.77 | 0.77 | 0.81 | 0.76 | 0.76 |
| Naïve Bayes | 0.79 | 0.78 | 0.77 | 0.80 | 0.78 | 0.77 | 0.81 | 0.78 | 0.77 | 0.81 | 0.78 | 0.77 |
| SVM | **0.85** | **0.85** | **0.85** | 0.83 | 0.82 | 0.82 | 0.82 | 0.80 | 0.80 | 0.82 | 0.80 | 0.80 |
| Logistic Regression | 0.83 | 0.82 | 0.81 | 0.81 | 0.78 | 0.77 | 0.81 | 0.76 | 0.76 | 0.80 | 0.75 | 0.74 |

## 5 Conclusion

In this paper, several classification methods for sentiment analysis are examined on political tweets obtained from active users. We exploited the sentiments keywords significant to major political parties in the tweets with respect to Delhi Elections 2020 and successfully annotated the corpus. The prepared annotated corpus from Twitter is used for political analysis. The data analysis significantly shows both, the supporting and opposing opinions of a substantial amount. The Support Vector Machine reports the highest precision, recall and F1 score among all the algorithms. The tweets collected at large undoubtedly represent a fraction of the actual population participating in the election process. Therefore, enormous attention is required for the sampling of data and generalization of models developed using the social media text. The performance of the models and the overall result can be enhanced by expanding the corpus with the incorporation of a large number of tweets and learning domain-specific keywords.